\documentclass[12pt,letterpaper]{article}

\newlength{\defbaselineskip}
\newcommand{\ket}[1]{|#1\rangle}
\newcommand{\bra}[1]{\langle#1|}

\newcommand{\half}{\frac{1}{2}}
\begin{document}
\title{Modal Interpretations and Relativity}
\author{Wayne C. Myrvold \\ Department of Philosophy \\
University of Western Ontario \\ London, ON Canada N6A 3K7 \\
e-mail: wmyrvold@uwo.ca}
\date{}

\maketitle

\begin{center}{To appear in \emph{Foundations of Physics
Letters.}}\end{center}
\bigskip
\par
%\doublespacing
\noindent A proof is given, at a greater level of
generality than previous `no-go' theorems, of the impossibility of
formulating a modal interpretation that exhibits `serious' Lorentz
invariance at the fundamental level.  Particular attention is
given to modal interpretations of the type proposed by Bub.

\bigskip
%\doublespacing
%\newpage
\section{Introduction} Modal interpretations of quantum mechanics posit that the
state vector obeys linear, unitary evolution at all times, and
supplement the state vector with a set of possessed properties
sufficiently rich to account for the occurrence of definite events
at the macroscopic level, including definite outcomes of
experiments, but sufficiently restricted so as to avoid a
Kochen-Specker contradiction.  The question arises whether this
can be done within the restrictions imposed by special relativity.
In a relativistic context, the notion of an instantaneous state of
a spatially extended system must be replaced by the notion of a
state on a spacelike hyperplane, or, more generally, a spacelike
hypersurface.  Since hyperplanes belonging to distinct foliations
will intersect, we must ask whether the definite properties
assigned to systems on these intersecting hyperplanes can be made
to mesh in a coherent way.

In connection with this question, two important `no-go' theorems
must be mentioned.  Dickson and Clifton \cite{DC} proved that the
answer is negative for a broad class of modal interpretations.
Berndl \emph{et al.} \cite{Berndl} showed that no theory that
shares with the Bohm theory the attribution of definite positions
at all times to particles can have the probability distributions
for these positions match the quantum-mechanical probability
distributions along every foliation.  The Dickson-Clifton proof
relies on an assumption concerning the transition probabilities
for possessed values, the assumption they call ``stability,'' but,
as Arntzenius \cite{Arnt} has pointed out, the stability
requirement is dispensable and the core of the proof concerns the
nonexistence of certain joint distributions yielding the
appropriate Born probabilities as marginals.  The proof in the
present paper is, in a sense, a generalization both of the proof
of Berndl \emph{et al.} and of Arntzenius' version of the
Dickson-Clifton proof.

Bub \cite{Bub92} introduced a class of modal interpretations that
single out some observable $R$ as having a definite value at all
times; this class includes the theories discussed by Berndl
\emph{et al.}, for which the preferred observable is position. As
Dickson and Clifton \cite[p. 36]{DC} point out, it is possible for
such an interpretation to evade their argument via a suitable
choice of preferred observable.  The existing `no-go' theorems,
therefore, leave it open whether a Bub-type modal interpretation
can be relativistically invariant. The question we want to ask is:
for a suitable choice of preferred observable $R$, can the
attribution of definite values to $R$ be made in such a way that
the probabilities concerning these definite values are given by
the Born-rule probabilities yield by the quantum-mechanical state
\emph{along every foliation}? As will be shown below, the answer
is negative, provided that the preferred properties are
\emph{local} properties and provided that certain transformations
of the quantum state are possible.  No assumptions about
transition probabilities for possessed values will be made.

\par {\quad} \par
\section{The proof}

Consider two systems, $S_i$, $i = 1,2$, which, during the times
that we are considering them, are localized (at least within the
approximations permitted by relativistic quantum field theory)
within regions that are large compared to their Compton
wavelengths but small compared to the distance between them.  We
do not assume that they are at rest with respect to each other.
Let $\alpha$ and $\beta$ be two hyperplanes of simultaneity for
some reference frame $\Sigma$. Let $p_i$ be a small region on
$\alpha$ in which the system $S_i$ is located, and let $q_i$ be a
region on $\beta$ in which $S_i$ located (see Figure 1). We assume
that the two systems are sufficiently far apart that $p_1$ is
spacelike separated from $q_2$, and $p_2$ is spacelike separated
from $q_1$. Let $\gamma$ be a spacelike hypersurface containing
$q_1$ and $p_2$, and let $\delta$ be a spacelike hypersurface
containing $p_1$ and $q_2$.

\begin{figure}[h]
\begin{picture}(360, 270)(-15,-10) \put(0,80){\line(1,0){340}} \put(0,180){\line(1,0){340}}
\put(-20,240){\line(1,0){400}} \put(-20,20){\line(1,0){400}}
\put(-20,20){\line(0,1){220}} \put(380,20){\line(0,1){220}}
 \put(10,50){\line(2,1){320}}
\put(10,210){\line(2,-1){320}} \put(69,50){\line(0,1){160}}
\put(71,50){\line(0,1){160}} \put(269,50){\line(0,1){160}}
\put(271,50){\line(0,1){160}} \put(68,35){$S_1$}
\put(268,35){$S_2$} \put(350,77){$\alpha$} \put(350,177){$\beta$}
\put(0,210){$\gamma$}\put(335,210){$\delta$} \put(78,70){$p_1$}
\put(255,70){$p_2$} \put(78,185){$q_1$} \put(255,185){$q_2$}
\put(-20,0){\textbf{Figure 1}. The hypersurfaces used in the
proof.}
\end{picture}
\end{figure}

%\begin{figure}[h]
%\framebox(400,220){

%\caption{}
%\end{figure}

%\begin{figure}[h]
%\begin{picture}(220, 120)
%\put(0,30){\line(1,0){204}} \put(0,90){\line(1,0){204}}
%\put(6,12){\line(2,1){192}} \put(10,108){\line(2,-1){192}}
%\put(41,12){\line(0,1){96}} \put(43,12){\line(0,1){96}}
%\put(161,12){\line(0,1){96}} \put(163,12){\line(0,1){96}}
%\put(40,0){$S_1$} \put(160,0){$S_2$} \put(210,28){$\alpha$}
%\put(210,88){$\beta$} \put(0,106){$\gamma$}\put(210,106){$\delta$}
%\end{picture}
%\caption{}
%\end{figure}

If $S_1$ and $S_2$ are isolated during the portion of their
evolution between $\alpha$ and $\beta$, or if the parts of their
environment with which they interact can be treated as effectively
classical and these interactions are local,  there will be unitary
operators $U_i$ such that the state of the combined system $S_1
\oplus S_2$ on $\beta$ will be related to its state on $\alpha$
by,
\begin{equation}\label{un}
\rho(\beta) = U_1 \otimes U_2 \; \rho(\alpha) \; U_1^{\dag}
\otimes U_2^{\dag}.
\end{equation}
If the regions $p_1$, $p_2$, $q_1$, $q_2$ are sufficiently small,
they may be treated as points, and we may regard $\gamma$ and
$\delta$ as hyperplanes of simultaneity for reference frames
$\Sigma'$, $\Sigma''$, respectively.  Let $\rho'(\gamma)$ be the
state according to $\Sigma'$ of the system $S_1 \oplus S_2$ at $t'
= t'_\gamma$, and let $\rho''(\delta)$ be the state according to
$\Sigma''$ at time $t'' = t''_\delta$.  We want to know how these
states are related to the $\Sigma$-states.

Someone using $\Sigma$ as a reference frame will judge that, if a
measurement of an observable $B_2$ is performed on $S_2$ at time
$t = t_\alpha$, and a measurement of an observable $A_1$ is
performed on $S_1$ at time $t = t_\beta$, the expectation value of
the product of the results of the measurements is
\begin{equation}
\mbox{Tr}[ \rho(\alpha)\; (U_1^{\dag} A_1 U_1  \otimes B_2) ].
\end{equation}
With respect to $\Sigma'$,  two such measurements occur
simultaneously, at $t' = t'_\gamma$.  The two reference frames
must agree on the probabilities of the outcomes of the
measurements.  The expectation value of the product of the two
measurements is, according to $\Sigma'$,
\begin{equation}
\mbox{Tr} [ \rho'(\gamma) \; A'_1 \otimes B'_2 ],
\end{equation}
where the operators $A'_1$, $B'_2$, are related to $A_1$, $B_2$
via the Lorentz transformation from $\Sigma$ to $\Sigma'$,
\begin{eqnarray}
\nonumber A'_1 &=& \Lambda_1 \: A_1 \: \Lambda_1^{\dag} \\
\nonumber B'_2 &=& \Lambda_2 \; B_2 \; \Lambda_2^{\dag} \\
 A'_1 \otimes B'_2 &=& (\Lambda_1 \otimes \Lambda_2) \;
(A_1 \otimes B_2) \;  (\Lambda_1^{\dag} \otimes \Lambda_2^{\dag})
 = \Lambda \; (A_1 \otimes B_2) \; \Lambda^{\dag}.
\end{eqnarray}
(Although the argument here does not depend on the Lorentz
transformation $\Lambda$ being a factorizable operator, it can be
proven [1] that it must, in fact, be factorizable.)

Since the two reference frames must agree on expectation values,
we must have:
\begin{equation}
\mbox{Tr}[ \rho(\alpha)\; (U_1^{\dag} A_1 U_1 \otimes B_2) ] =
 \mbox{Tr} [ \rho'(\gamma) \; A'_1 \otimes B'_2]
\end{equation}
A bit of algebraic manipulation yields,
\begin{equation}
\mbox{Tr} [(U_1 \otimes I_2)\:  \rho(\alpha) \: (U_1^{\dag}
\otimes I_2)\;(A_1 \otimes B_2)] = \mbox{Tr} [\Lambda^{\dag} \:
\rho'(\gamma) \: \Lambda \; (A_1 \otimes B_2) ]
\end{equation}
Since this must hold for arbitrary $A_1$, $B_2$, we must have,
\begin{equation}
(U_1 \otimes I_2)\;  \rho(\alpha) \; (U_1^{\dag} \otimes I_2) =
\Lambda^{\dag} \: \rho'(\gamma) \: \Lambda,
\end{equation}
or,
\begin{equation}
\rho'(\gamma) = \Lambda \; (U_1 \otimes I_2)\;  \rho(\alpha) \;
(U_1^{\dag} \otimes I_2) \; \Lambda^{\dag}.
\end{equation}
 Similarly,
\begin{equation}
\rho''(\delta)  = \Lambda'  \; (I_1 \otimes U_2) \; \rho(\alpha)
\; (I_1 \otimes U_2^{\dag}) \; {\Lambda'}^{\dag},
\end{equation}
where $\Lambda' = \Lambda'_1 \otimes \Lambda'_2$ is the
transformation from $\Sigma$ to $\Sigma''$.

Now, the Lorentz boost operators $\Lambda$, $\Lambda'$ merely
effect a transformation from a state given with respect to one
reference frame's coordinates to  one given with respect to
another reference frame's coordinates. In what follows, it will be
more convenient to utilize the coordinate basis of one reference
frame, $\Sigma$, for all states, even those on hypersurfaces that
are not equal-time hyperplanes for $\Sigma$.   We will therefore
transform the states $\rho'(\gamma)$ and $\rho''(\delta)$ back
into $\Sigma$'s coordinate basis,
\begin{equation}
\rho(\gamma) = \Lambda^\dag \; \rho'(\gamma) \; \Lambda = U_1
\otimes I_2 \; \rho(\alpha) \ U_1^{\dag} \otimes I_2
\end{equation}
\begin{equation}
\rho(\delta) = {\Lambda'}^\dag \; \rho''(\delta) \; \Lambda' = I_1
\otimes U_2 \; \rho(\alpha) \; I_1 \otimes U_2^{\dag}
\end{equation}

For more general interactions of the system $S_1 \oplus S_2$
 with its environment, the evolution of reduced state of
the system will, provided that these interactions are local
interactions, have a  Kraus representation \cite{Kraus} consisting
of factorizable operators (see \cite{CH} for a discussion):
\begin{equation}\label{non}
\rho(\beta) = \sum_{m,n} K_{1m} \otimes K_{2n} \;\; \rho(\alpha)
\; \; K_{1m}^{\dag} \otimes K_{2n}^{\dag},
\end{equation}
where
\begin{equation}
\sum_k K_{i k}^{\dag} K_{i k} = I_i.
\end{equation}

The corresponding states on $\gamma$ and $\delta$ are given by,
\begin{eqnarray}
\rho(\gamma) &=& \sum_m K_{1m} \otimes I_2 \;\; \rho(\alpha) \; \;
K_{1m}^{\dag} \otimes I_2,
\\
\rho(\delta) &=& \sum_n I_1 \otimes K_{2n} \;\; \rho(\alpha) \; \;
I_1 \otimes K_{2n}^{\dag}.
\end{eqnarray}

Suppose that $A_1$ and $A_2$ are definite properties of $S_1$ and
$S_2$, respectively, on $\alpha$, and $B_1$ and $B_2$ are definite
properties on $\beta$.  If these are \emph{local}
properties---that is, properties possessed by the system
irrespective of considerations of the rest of the universe ---
then the value of $A_1$ possessed by $S_1$ at $p_1$ is possessed
by it without reference to the hypersurface containing $p_1$ being
considered, and similarly for the other points of intersection
$p_2$, $q_1$, $q_2$.  (Indicating a particular outcome is,
presumably, a local property of apparatus pointers. Being 100 km
from New York City is not.)

We will require that the probability distributions for possessed
values of local properties satisfy:

\begin{quote}
\emph{Relativistic Born Rule.} For any spacelike hypersurface
$\sigma$, if the quantum state of the combined system $S_1 \oplus
S_2$ on $\sigma$ is $\rho(\sigma)$, and if $X_1$ and  $Y_2$ are
local definite properties of $S_1$ and $S_2$ on $\sigma$, then the
probability that $X_1 = x$ and $Y_2 = y$ on $\sigma$ is equal to
$\mbox{Tr}[P_{X_1}(x) \: P_{Y_2}(y)\: \rho(\sigma)]$, where
$P_{X_1}(x)$ and $P_{Y_2}(y)$ are the projections onto the
eigenspaces $X_1 = x$ and $Y_2 = y$, respectively.
\end{quote}

Even if our modal interpretation is agnostic about transition
probabilities, if the probabilities regarding the possessed values
of the definite observables are to satisfy the Born rule on all
four hypersurfaces, it must be possible for there to be a joint
probability distribution over all four of our observables, that
yields as marginals the Born probabilities on all four
hyperplanes.  Suppose, then, that there is such a distribution,
Pr($a_{1i}$, $a_{2j}$, $b_{1k}$, $b_{2l})$, this being the
probability that $S_1$ has $A_1 = a_{1i}$ at $p_1$, $S_2$ has $A_2
= a_{2j}$ at $p_2$, $S_1$ has $B_1 = b_{1k}$ at $q_1$, and $S_2$
has $B_2 = b_{2l}$ at $q_2$.  We will make no assumption about
this joint distribution other than that it yield the Born rule
probabilities as marginals on all four hypersurfaces, $\alpha$,
$\beta$, $\gamma$, $\delta$,
\begin{eqnarray}
\nonumber \sum_{k, \: l} \mbox{Pr}(a_{1i}, \: a_{2j}, \: b_{1k},\:
b_{2l}) = \mbox{Tr}[ P_{A_1}(a_{1i})\: P_{A_2}(a_{2j}) \:
\rho(\alpha) ]
\\
\nonumber \sum_{i,j}\mbox{Pr}(a_{1i}, \: a_{2j}, \: b_{1k},\:
b_{2l})= \mbox{Tr}[ P_{B_1}(b_{1k})\: P_{B_2}(b_{2l}) \:
\rho(\beta) ]
\\
\nonumber \sum_{i, l}\mbox{Pr}(a_{1i}, \: a_{2j}, \: b_{1k},\:
b_{2l}) = \mbox{Tr}[ P_{B_1}(b_{1k})\: P_{A_2}(a_{2j}) \:
\rho(\gamma) ]
\\
\sum_{j, k}\mbox{Pr}(a_{1i}, \: a_{2j}, \: b_{1k},\: b_{2l}) =
\mbox{Tr}[ P_{A_1}(a_{1i})\: P_{B_2}(b_{2l}) \: \rho(\delta) ].
\end{eqnarray}

Because of the relations between the states on the hyperplanes
considered, the existence of such a joint distribution is
equivalent to the existence of a joint distribution yielding, as
marginals, the statistics in state $\rho(\alpha)$ for the
observables $A_1 \otimes A_2$,  $A_1 \otimes C_2$, $C_1 \otimes
A_2$, $C_1 \otimes C_2$, where
\begin{equation}
C_i = U_i^{\dag} \: B_i \: U_i
\end{equation}
in the case of unitary evolution (\ref{un}); in the case of
non-unitary evolution (\ref{non}), $C_i$ is the `mixed
observable,'
\begin{equation}
C_i = \sum_k K_{i k}^{\dag} \; B_i \; K_{i k}.
\end{equation}

It has long been recognized \cite{Fine} that violation of a Bell
inequality entails the nonexistence of such a joint distribution.
If $\rho(\alpha$) is a state such that a Bell inequality can be
derived for the observables $A_1$, $C_1$, $A_2$, $C_2$, then,
assuming the relativistic Born rule, it cannot be the case that
$A_1$ is definite at $p_1$, $A_2$ is definite at $p_2$, $B_1$ is
definite at $q_1$, and $B_2$ is definite at $q_2$.

Let us now apply these considerations to Bub's modal
interpretation, which selects some observable $R$ as
always-definite. Let $R_1$, $R_2$ be always-definite observables
of $S_1$ and $S_2$, respectively, such that the possession of any
definite value of these observables is a local property of the
system possessing it. We will assume that each $R_i$ has at least
two distinct eigenvalues, $\{r_i^+ , r_i^- \}$. Let $\{
\ket{r_i^+},\ket{r_i^-} \}$ be corresponding eigenstates.

Suppose, now, that the system is prepared so as to be, on
$\alpha$, in the Hardy-Jordan state \cite{HJ},
\begin{equation}\label{Halpha}
\ket{\psi(\alpha)} =  \frac{1}{2 \sqrt{3}} \left(
\ket{r_1^+}\ket{r_2^+} - \ket{r_1^+}\ket{r_2^-}-
\ket{r_1^-}\ket{r_2^+}  - 3 \: \ket{r_1^-}\ket{r_2^-} \right).
\end{equation}
Let us also assume that it is possible to effect a Hadamard
transformation of the $R$-eigenstates,
\begin{eqnarray}\label{had}
U_i \: \ket{r_i^+} &=& \frac{1}{\sqrt{2}} \left(\ket{r_i^+} +
\ket{r_i^-} \right) \nonumber
\\
U_i \: \ket{r_i^-} &=& \frac{1}{\sqrt{2}} \left(\ket{r_i^+} -
\ket{r_i^-} \right).
\end{eqnarray}
%Such a transformation will occur if, for example, the Hamiltonian
%of the system is of the form
%\begin{equation}
%H_i = i \hbar \: \omega \left(\ket{r_i -}_i \bra{r_i +}_i -
%\ket{r_i +}_i \bra{r_i -}_i \right),
%\end{equation}
%and the system is allowed to evolve undisturbed for a time equal
%to $\tau = \pi / 4 \omega$.

Between $\alpha$ and $\beta$, we apply a Hadamard transformation
to each system separately. The state on $\beta$ of the combined
system will then be given by
\begin{equation}\label{Hbeta}
\ket{\psi(\beta)} = \frac{1}{\sqrt{3}} \left( \ket{r_1^+}
\ket{r_2^-} + \ket{r_1^-} \ket{r_2^+} - \ket{r_1^+} \ket{r_2^+}
\right).
\end{equation}
The state on $\gamma$  is
\begin{eqnarray}\label{Hgamma}
\ket{\psi(\gamma)} &=& U_1 \otimes I_2 \:  \ket{\psi(\alpha)}
\nonumber
\\
&=& \frac{1}{\sqrt{6}} \left( \ket{r_1^-} \ket{r_2^+} +
\ket{r_1^-} \ket{r_2^-} - 2 \: \ket{r_1^+} \ket{r_2^-} \right).
\end{eqnarray}
The state on $\delta$ is
\begin{eqnarray}\label{Hdelta}
\ket{\psi(\delta)} &=& I_1 \otimes U_2 \:  \ket{\psi(\alpha)}
\nonumber
\\
&=& \frac{1}{\sqrt{6}} \left( \ket{r_1^+} \ket{r_2^-} +
\ket{r_1^-} \ket{r_2^-} - 2 \: \ket{r_1^-} \ket{r_2^+} \right).
\end{eqnarray}

Suppose that, on  $\alpha$, $R_1$ and $R_2$ have  the values
$(r_1^+$, $r_2^+)$.   Since $R_1$ is, by assumption, a local
property of $S_1$, $S_1$ must have the same value $R_1 = r_1^+$ on
the hypersurface $\delta$.  The state (\ref{Hdelta}) assigns
probability zero to the pair of values $( r_1^+, r_2^+ )$, and so,
on $\delta$, $R_2$ must, with probability one, have the value
$r_2^-$. Since $R_2$ is a local property of $S_2$, $R_2$ has the
value $r_2^-$ on $\beta$ as well.  A parallel argument leads to
the conclusion that, if $R_2$ has the value $r_2^+$ on $\alpha$,
$R_1$ has the value $r_1^-$ on $\beta$.

We therefore conclude that, if $R_1$ and $R_2$ have the values $(
r_1^+, r_2^+ )$ on $\alpha$, they have the values $( r_1^-, r_2^-
)$ on $\beta$.  But, whereas $( r_1^+, r_2^+ )$ has probability
$1/12$ on $\alpha$, inspection of (\ref{Hbeta}) shows that $(
r_1^-, r_2^- )$ has probability zero on $\beta$. Therefore, it is
impossible to satisfy the Born-rule probabilities for possessed
values of $R_1$ and $R_2$ on all four of the hypersurfaces
$\alpha$, $\beta$, $\gamma$, $\delta$.

The above argument, as it stands, does not apply to those modal
interpretations that use the Schmidt biorthogonal decomposition of
the state to pick out the preferred observables.  The argument can
be made to apply with a simple modification.  Associate with each
of the systems $S_i$ a second system $A_i$, among whose
observables is a `pointer' observable with eigenstates
$\ket{p_i^\pm}_{A_i}$ that can be made to interact with $S_i$ in
such a way that the values of the pointer observables become
correlated with the values of $R_i$.  Take the state of the system
on $\alpha$ to be the state obtained from (\ref{Halpha}) by
replacing $\ket{r_i^ \pm}$ by $\ket{r_i^ \pm }_{S_i} \ket{p_i ^\pm
}_{A_i}$. It is easy to check that the orthogonal decomposition of
the reduced  density operator for $S_i$ is nondegenerate on all
four hypersurfaces and yields $R_i$ as definite properties on
these hypersurfaces. The argument requires that we apply a
Hadamard transformation to the combined system-apparatus state,
\begin{eqnarray}\label{had2}
U_i \:  \ket{r_i^+ }_{S_i} \ket{p_i^+ }_{A_i} &=&
\frac{1}{\sqrt{2}} \left(\ket{r_i^+ }_{S_i} \ket{p_i^+ }_{A_i} +
\ket{r_i^- }_{S_i} \ket{p_i^- }_{A_i} \right) \nonumber
\\
U_i \: \ket{r_i^- }_{S_i} \ket{p_i^- }_{A_i} &=&
\frac{1}{\sqrt{2}} \left(\ket{r_i^+ }_{S_i} \ket{p_i^+ }_{A_i} -
\ket{r_i^- }_{S_i} \ket{p_i^- }_{A_i} \right).
\end{eqnarray}

\par {\quad} \par
\section{Idealizations relaxed}

The above argument presumes that it is possible to keep the system
isolated while performing a Hadamard transformation; this must be
regarded as somewhat of an idealization, as no system is ever
completely isolated from its environment. Bub \cite[\S 5.2]{Bub98}
has argued that the preferred observable should be stable with
respect to environmentally induced decoherence.  If this is the
case, such decoherence will tend to turn coherent superpositions
of distinct $R$-values into improper mixtures. Because of this the
transformation invoked in the preceding section, which mixes
distinct $R_i$-eigenspaces, may in practice be tremendously
difficult. The issues with which we are concerned are, however,
matters of principle; a theory that permits violations of the
relativistic Born rule is not a relativistic theory even if
situations that mandate such a violation are difficult to achieve
in practice and the natural occurrence of such situations is
extremely improbable. One might contemplate the possibility,
however, of there being a limit in principle to the extent to
which the system can be isolated from its environment; the
always-definite observable might, for example, interact with the
vacuum fields. We should, therefore, ask whether a version of
argument can survive such an ineliminable environmental
interaction. We will still require that the relativistic Born rule
be satisfied for arbitrary initial states, but will no longer
assume that the system can be regarded as isolated while a
Hadamard transformation is performed.

Suppose that we apply to $S_i$ an external potential $H_i$. If
$H_i$ is much larger than the interaction of the system with its
environment, then the evolution of the system will, for
sufficiently short periods of time, be dominated by this term and
will approximate the evolution that would obtain if there were no
environmentally induced decoherence.  It is therefore worth
pointing out that a full Hadamard transformation is not necessary
for a violation of the relativistic Born rule, and that this can
be achieved, for a suitable initial state, by an arbitrarily small
rotation of the state.  To show this, we consider, not the
Hardy-Jordan state, but the singlet state,
\begin{equation}
\ket{\psi(\alpha)} = \frac{1}{\sqrt{2}} \left( \ket{r_1^+}
\ket{r_2^-} - \ket{r_1^-} \ket{r_2^+} \right).
\end{equation}
Apply to the systems $S_1$ and $S_2$ potentials whose effect is to
rotate the states in opposite directions:
\begin{eqnarray}
H_1 &=& i \hbar \omega \left( \ket{r_1^-}\bra{r_1^+} -
\ket{r_1^+}\bra{r_1^-} \right)
\\
H_2 &=& - i \hbar \omega \left( \ket{r_2^-}\bra{r_2^+} -
\ket{r_2^+}\bra{r_2^-} \right)
\end{eqnarray}
Take the time interval $\Delta t$ between $\alpha$ and $\beta$ to
be sufficiently small that the effects of environmentally induced
decoherence are negligible.  We will then have the states on our
other hypersurfaces given approximately by
\begin{eqnarray}
 \nonumber \ket{\psi(\gamma)} = \frac{1}{\sqrt{2}} \left( \sin
 \phi\;
\ket{r_1^+}\ket{r_2^+} \right. + \cos \phi \; \ket{r_1^+}
\ket{r_2^-}
\\  - \left. \cos \phi\; \ket{r_1^-} \ket{r_2^+} + \sin
\phi\; \ket{r_1^-} \ket{r_2^-} \right)
\end{eqnarray}
\begin{eqnarray}
 \nonumber \ket{\psi(\delta)} = \frac{1}{\sqrt{2}} \left( \sin
 \phi \;
\ket{r_1^+} \ket{r_2^+} + \cos \phi \; \ket{r_1^+} \ket{r_2^-}
\right .
\\\left. - \cos \phi \; \ket{r_1^-} \ket{r_2^+} + \sin \phi \;
\ket{r_1^-} \ket{r_2^-} \right),
\end{eqnarray}
\begin{eqnarray}
\nonumber \ket{\psi(\beta)} = \frac{1}{\sqrt{2}} \left( \sin 2
\phi \; \ket{r_1^+} \ket{r_2^+} + \cos 2 \phi \; \ket{r_1^+}
\ket{r_2^-} \right.
\\
\left. - \cos 2 \phi \; \ket{r_1^-} \ket{r_2^+} + \sin 2 \phi \;
\ket{r_1^-} \ket{r_2^-} \right),
\end{eqnarray}
where $\phi = \omega \Delta t$.

Let $R_i^{+}(x)$ be the proposition
that $R_i$ has value $r_i^+$ at spacetime point $x$, and similarly
for $R_i^{-}(x)$ .  If there is a joint distribution over the
possessed values of $R_1$ and $R_2$ on $\alpha$ and $\beta$, then
we should have
\begin{eqnarray}
\nonumber 0 &\leq& \mbox{Pr}[R_1^+(p_1) \:  \& \: R_2^-(q_2)] +
\mbox{Pr}[R_1^-(q_1) \: \& \: R_2^+(p_2)] \\ &+&
\mbox{Pr}[R_1^+(q_1) \: \& \: R_2^+(q_2)] - \mbox{Pr}[R_1^+(p_1)
\: \& \: R_2^+(p_2)] \; \leq \; 1.
\end{eqnarray}

Assuming that these probabilities are given by the Born rule, in
our example this amounts to
\begin{equation}
0 \; \leq \; \cos^2 \phi + \half \sin^2 2 \phi \; \leq \; 1.
\end{equation}
This is violated for $0 < |\phi| < \pi/4$, and hence for
arbitrarily small $\phi$.

\par {\quad} \par
\section{Lorentz invariance, serious and otherwise}

``\emph{Zur Elektrodynamik bewegter K\"orper}'' \cite{AE} opens
with the observation that electrodynamics, as it was understood at
the time, leads to asymmetries in the theoretical description that
are not present in the phenomena, in that the theoretical
description distinguishes between bodies in motion and those at
rest, in spite of the fact that the observable phenomena depend
only on the \emph{relative} motion of bodies.  Such
considerations, says Einstein, suggest that there is in fact
nothing corresponding to absolute rest.  He goes on in the paper
to show how to reconcile electrodynamics with this suggestion; to
do so involves rejecting the notion also that there is anything
corresponding to absolute simultaneity of spatially separated
events. The transformation between inertial coordinates, as
measured by physical rods and clocks, must be given by the Lorentz
transformation.

Now, it is certainly possible to suppose that there is a
distinguished state of absolute rest; provided that this state is
defined with respect to the matter in the Universe or some other
physical structure, it is even possible for a theory that posits
such a state to do this while preserving Lorentz invariance of the
formulas of the theory. Similarly, a theory may introduce a
preferred foliation in a Lorentz invariant manner. To do so,
however, is to ignore the reasons why we should be interested in
Lorentz invariance in the first place. The observable phenomena
pick out neither a preferred rest frame nor a preferred relation
of distant simultaneity. This is precisely what is to be expected
if there is in reality no preferred state of rest and no
distinguished relation of distant simultaneity, and so we
hypothesize that this is, in fact, the case, and impose Lorentz
invariance to ensure that an assumption of a preferred Lorentz
frame is not concealed in our choice of coordinates. To introduce
a preferred foliation in a Lorentz invariant manner is to abandon
what Bell \cite[p. 180]{Bell} calls ``serious Lorentz
invariance.''

The ``Lorentz-Covariant modal scheme'' outlined by Dieks
\cite{Dieks} evades the Dickson-Clifton proof by rejecting the
Dickson-Clifton stability condition; it also evades the Arntzenius
version of that proof, and the proof of the present paper, by
rejecting the relativistic Born rule; on this scheme, the
Born-rule probabilities do not give the probabilities for
possessed values at all times for all foliations.  Similarly,
D\"urr \emph{et al.} \cite{Durr} produce a covariant trajectory
model by introducing, as part of the dynamical structure of the
theory, a foliation with respect to which the ``quantum
equilibrium'' condition $P = |\psi|^2$ is satisfied. As nothing in
the observable phenomena depend on the particular choice of such a
foliation, the distinguished foliation introduces into the
theoretical description an asymmetry not present in the phenomena.
The reasons for rejecting such a move, therefore, are precisely
the same as the reasons for Einstein's dissatisfaction with a
formulation of electrodynamics that invokes a preferred rest
frame.

As Bell points out, we do not have a precise criterion for
seriousness of Lorentz Invariance.  It seems clear, however, that
the relativistic Born rule should be satisfied by any
interpretation of quantum mechanics with a claim to serious
Lorentz invariance.

\textbf{Acknowledgment.} I would like to thank two anonymous
referees for their helpful suggestions.

%-------------------------Bibliography-------------------------------------
\newpage
%\noindent {\Large {\textbf{References}} }

 \end{document}